\documentclass{article}

\usepackage[preprint]{neurips_2022}

\usepackage[labelfont=bf]{caption}
\usepackage{multirow}
\usepackage[utf8]{inputenc} 
\usepackage[T1]{fontenc}    
\usepackage{hyperref}       
\usepackage{url}            
\usepackage{booktabs}       
\usepackage{amsfonts}       
\usepackage{nicefrac}       
\usepackage{microtype}      
\usepackage{xcolor}         

\usepackage{graphicx}
\usepackage{caption}
\title{Multi-Prompt Fine-Tuning of Foundation Models for Enhanced Medical Image Segmentation}
\author{
Xiangru Li \\
  Emory University \\
  \texttt{xiangru.li@emory.edu} \\
  \And
  Yifei Zhang \\
  Emory University \\
  \texttt{yifei.zhang2@emory.edu} \\
  \And
  Liang Zhao \\
  Emory University \\
  \texttt{liang.zhao@emory.edu} \\
}
\begin{document}

\maketitle
\begin{abstract}
The Segment Anything Model (SAM) is a powerful foundation model that introduced revolutionary advancements in natural image segmentation. However, its performance remains sub-optimal when delineating the intricate structure of biomedical images, where multiple organs and tissues intertwine in a single image. In this study, we introduce a novel fine-tuning framework that leverages SAM’s ability to bundle and process multiple prompts per image and seeks to improve SAM’s performance in medical images. We first curated a medical image dataset that consists of CT scans of lesions in various organs, each with two annotations for organs and lesions respectively. Then, we fine-tuned SAM's mask decoder within our framework by batching both bounding boxes generated from ground truth masks as reference. The batched prompt strategy we introduced not only addresses the inherent complexity and ambiguity often found in medical images but also substantially enhances performance metrics when applied onto a wide range of segmentation tasks.


\end{abstract}

\section{Introduction}

Recent advances in computer vision foundation models have bolstered their application in segmentation tasks \cite{ANAYAISAZA2021100723,7444155, Sharma2010-lk}, exemplified by the achievements of the Segment Anything (SA) project. The Segment Anything Model (SAM), introduced by the SA project, enjoys robust zero-shot performance comparable to many supervised methods \cite{kirillov2023segment}. However, its effectiveness fluctuates when applied to medical images with different modalities and amorphous structural boundaries \cite{Mazurowski_2023,huang2023segment,chen2023masam}.

Prior research like MedSAM~\cite{ma2023segment} has created a foundation model for universal medical image segmentation with promising results. However, its training framework involved only a single prompt per image, which failed to utilize SAM's prompt encoder's potential to take multiple prompts per image - a feature known to reduce segmentation ambiguity in both natural and medical images \cite{kirillov2023segment,Mazurowski_2023}. Given the intricate structures present in medical images and the wealth of multi-label scans available within medical databases, such as organ-lesion pairs and multi-organ annotations \cite{gibson_eli_2018_1169361,JimenezdelToro2016,Rister2020}, we introduce a framework that allows for the fine-tuning of SAM using multiple ground truth masks per image. We retrieved the benchmark dataset from the Medical Segmentation Decathlon (MSD) \cite{Antonelli_2022} to evaluate the performance of our framework. Each image in the dataset contains two expert-annotated ground truth masks highlighting the locations of organs and lesions. We produced and batched bounding box prompts using those annotations to fine-tune SAM, enabling the model to learn from the position encoding of both structures. Our results highlights the powerful synergy between prompts, achieving enhanced segmentation accuracy as compared to single-prompt training methods and foundation models in both lesion and organ segmentation tasks. These results demonstrate the potential of our work to streamline the medical image segmentation process by presenting a framework for fine-tuning foundation models with great efficiency, thereby achieving robust, high-accuracy results across diverse medical imaging modalities.
\vspace{-7pt}
\section{Method}
\vspace{-7pt}
Currently, SAM supports three segmentation modes: automatic segmentation, prompted segmentation using points, and prompted segmentation using bounding boxes. Previous research on the application of SAM in medical images has concluded that the bounding box approach is the most efficient and accurate method for delineating complex structures in medical images \cite{huang2023segment,ma2023segment,Mazurowski_2023}. In our approach, we fine-tune the model by combining bounding box prompts of multiple ground truth masks in the same image. By incorporating the positional encoding of multiple regions of interest (ROI) when generating segmentation masks, our approach effectively addresses the inherent complexity in medical images, which substantially mitigates prompt ambiguity and enhances segmentation accuracy.

\begin{figure}[h!]
    \centering
    \includegraphics[width=\linewidth]{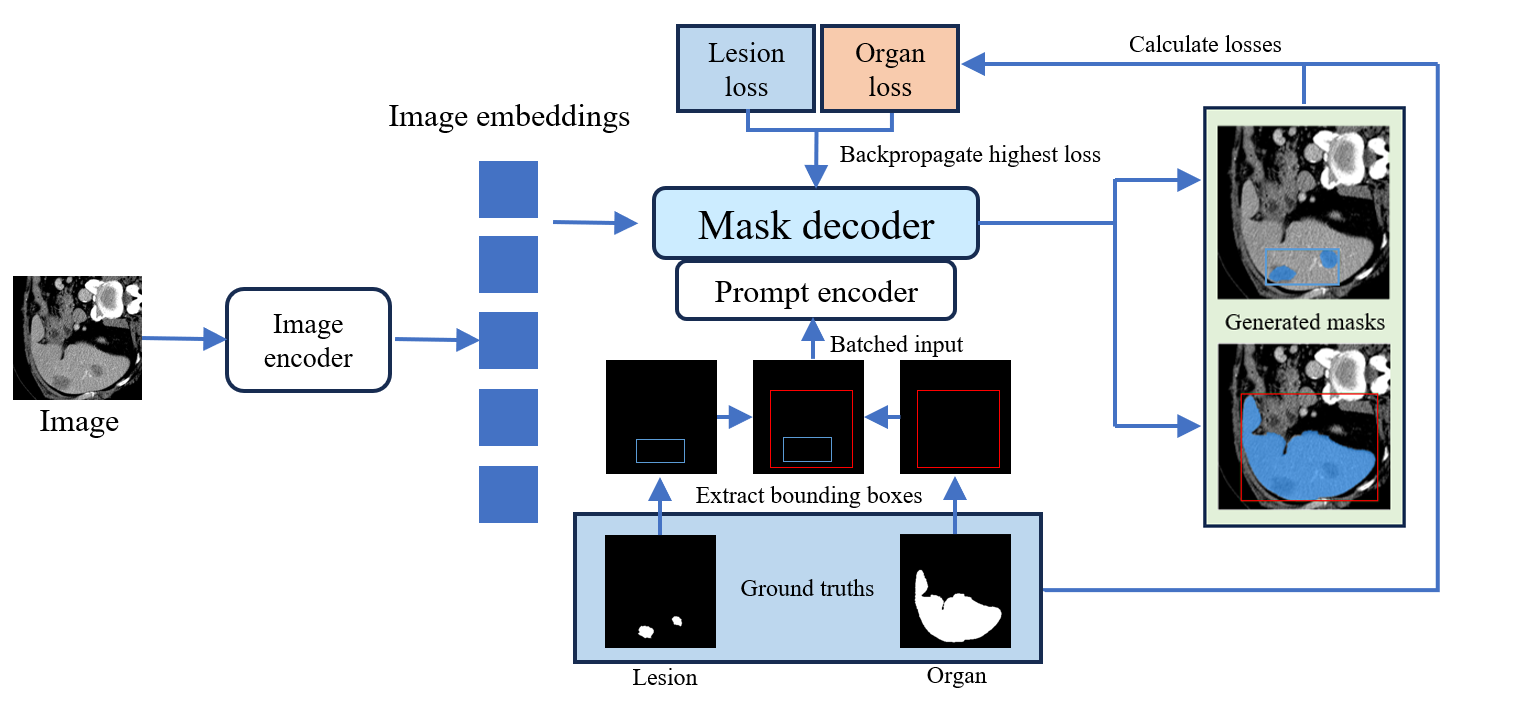}
    \caption{Framework for fine-tuning SAM for medical image segmentation. Multiple bounding box prompts per image embedding are batched by the prompt encoder and passed to the mask decoder.}
    \label{fig:framework}
\end{figure}

Our proposed fine-tuning framework, which we refer to as co-training, is illustrated in Figure \ref{fig:framework}. We begin by initializing our model with the pre-trained SAM ViT-Base model. We retain the image encoder from SAM to generate image embeddings before training to reduce computational burden \cite{ma2023segment}. Subsequently, our framework can take multiple masks and as exemplified in Figure \ref{fig:framework}, we employed two ground truth masks, one for the organ and the other for the lesion. These masks are strongly correlated to each other. For instance, the supervision from the organ mask can not only help organ segmentation, but also aid to reinforce the location of lesions, and vice-versa. These masks are used to produce two bounding boxes, pinpointing the ROIs within the image. To simulate possible human errors in clinical application, a perturbation is applied to the bounding boxes to fluctuate their dimensions within a certain range, which are hyperparameters based on the ROI type and size to enhance the model's robustness. After merging the perturbed bounding boxes, the prompt encoder processes it to form positional encoding, which, alongside image embeddings, is passed to the mask decoder to produce two segmentation masks corresponding to each prompt. In each epoch, the loss is calculated by performing an unweighted sum between Dice loss and cross-entropy loss, a method proven to be robust in segmentation tasks \cite{ma2023segment,Isensee2021-rr}. We compute the loss of each generated mask separately with their respective ground truth masks. After that, only the highest loss will be used to calculate the gradient to ensure proper optimization on both annotations. 
\vspace{-10pt}
\vspace{-7px}
\section{Results}
\vspace{-7px}
To investigate the feasibility of our proposed framework, we curated two benchmark datasets from the Medical Segmentation Decathlon (MSD) dataset \cite{Antonelli_2022}. These datasets encompass CT scans for both liver and pancreatic lesions, each labeled with corresponding organ and lesion annotations by radiologists. We preprocessed the 3D scans into 2D images by performing random slices on the z-axis, yielding 98 liver scans and 281 pancreatic scans in total \cite{gu2023essa}. Each scan displayed unbalanced labels between larger structures (organs) and smaller ROIs (lesions), a challenge often encountered in medical segmentation tasks. We then partitioned the datasets, allocating 70\% for training, 15\% for validation, and 15\% for testing. To accommodate the size disparity between structures, specific bounding box perturbation ranges have been defined for organ and lesion masks, respectively. Each range is uniformly maintained across all experimental groups to ensure the validity of our results. While testing, we utilized various evaluation metrics such as Intersection over Union (IoU), Dice Similarity Coefficient (DSC), and Normalized Surface Distance (NSD). We compared the co-trained model's lesion segmentation results with a model fine-tuned solely on lesion ground truths and used MedSAM as a baseline. We further investigated the adaptability of our co-trained model by applying the same model to organ segmentation within the same dataset, comparing its performance against a model fine-tuned on organ masks and the baseline. Detailed segmentation performance metrics are presented in Table \ref{tab:results}. In each task and metric, the best results are highlighted with boldface font.
\vspace{-5pt}
\begin{table*}[h!]
  
  \centering
  \caption{Lesion and organ segmentation results of the co-trained model, single-prompt fine-tuned model and MedSAM as the baseline, measured by IoU, DSC, and NSD with a threshold of 1 pixel.}
  \label{tab:results}
   \resizebox{\linewidth}{!}{%
  \begin{tabular}{c|c|ccc|ccc}
    \toprule
    Dataset & Method   &  & Lesion Segmentation &     &   &  Organ Segmentation & \\
    \hline
    \multirow{4}{*}{Pancreas Dataset}

        &  & IoU (\%)& DSC (\%)& NSD & IoU (\%)& DSC (\%) & NSD
    
        \\
        \cline{2-8}
        & MedSAM  & 50.12& 63.41 & 7.760 & 84.02& 90.61& 9.940\\
        
        & Single Prompt  & 62.02& 73.27 &4.854 & 72.44& 82.59 & 15.18
        
        \\
        & Co-train  & \textbf{65.74}&\textbf{ 77.98}&\textbf{ 3.981}&\textbf{ 85.41}&\textbf{ 91.93}&\textbf{ 9.170}
        \\

    \hline
    \multirow{3}{*}{Liver Dataset}
      & MedSAM  & 65.13& 77.87 & 4.079 & 60.01 & 73.04& 8.816
        \\
       
        & Single Prompt &72.61& 83.69& 3.003 & 74.98 & 85.10 & 4.392 
        \\
         & Co-train &\textbf{ 77.74 }&\textbf{ 87.16} & \textbf{2.258} &\textbf{ 77.20} & \textbf{86.72 } & \textbf{4.031}
        \\

    \bottomrule
  \end{tabular}
   }
   
\end{table*}

From the table, we can see a co-trained model outperforms all other comparison methods. Specifically, in lesion segmentation tasks, co-trained models consistently yield the best performance on all metrics in both datasets, with 5-31\%, 4-23\%, and 2-48\% improvement in terms of IoU, DSC, and NSD scores, respectively, compared with the baseline and single-prompt tuned models. The outstanding performance continues in organ segmentation tasks. Our co-trained model shows a 3-28\%, 1-19\%, and 2-51\% enhancement in performance compared to other groups in terms of IoU, DSC, and NSD scores, respectively. These results underscore the potential of our framework to produce outstanding models that exhibit not only high segmentation accuracy as compared to traditional methods but also robust generalization properties applicable to various structures involved in the training process, which has the potential to boost the efficiency of future multi-target segmentation endeavors.

\begin{figure}[h!]
    \centering
    \includegraphics[width=0.8\columnwidth]{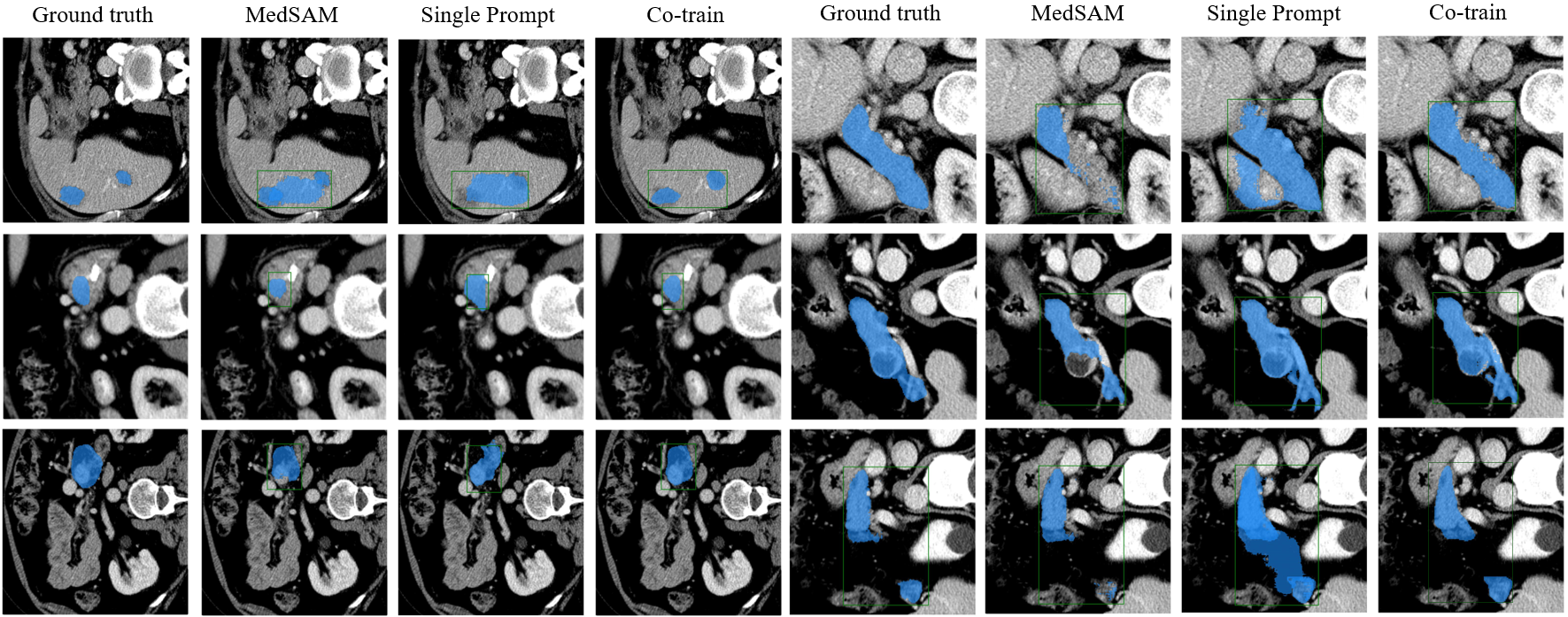}
    \caption{Visualizations of Results from Different 2D Medical Image Segmentation Models: Co-trained Models, Single-Prompt Trained Models, and MedSAM.}
    \label{fig:resimg}
\end{figure}

To visualize those metrics, we present some of our segmentation results in Figure \ref{fig:resimg}, side-by-side with the ground truth, MedSAM, and single-prompt tuned models. Overall, the segmentations produced by co-trained models most closely mirror the ground truths. Our results also demonstrate minimal segmentation ambiguity, as evidenced by the distinct boundaries compared to some other results, which display varying degrees overfitting and underfitting with ambiguous boundaries.

\section{Conclusion}

In this study, we introduce a novel fine-tuning framework, designed to harness the multi-prompt capabilities of the Segment Anything Model (SAM), for advanced medical image segmentation. Notable increases in various performance metrics underscore the potential of our framework as a robust and efficient solution to challenges in medical image segmentation.

\label{headings}

\bibliography{refs}
\bibliographystyle{plain}
\section*{Potential Negative Societal Impacts}
Despite the promising advancements demonstrated by our framework in medical image segmentation, potential negative societal impacts warrant consideration. The integration of automated segmentation models into clinical workflows could inadvertently precipitate over-reliance on machine learning models, possibly leading to oversight and reduced diligence in manual review. The error propagation from automated segmentation could potentially contribute to inaccurate diagnosis and suboptimal treatment planning, impacting patient outcomes adversely. It’s crucial for clinicians to maintain an active role in reviewing and verifying automated segmentation results to mitigate such risks. Additionally, the generalization of our model to diverse and underrepresented patient populations remains a challenge, as biases in the training data could potentially perpetuate health disparities. Vigilant assessment and continuous refinement of models, in conjunction with comprehensive and diverse datasets, are imperative to ensure equitable and reliable performance across diverse patient demographics.
\end{document}